\documentclass{article}
\newcommand{\be}{\begin{equation}}
\newcommand{\ee}{\end{equation}}
\newcommand{\nbar}[1]{\overline{#1}}                       

\def\bea{\begin{eqnarray}}
\def\eea{\end{eqnarray}}
\def\beas{\begin{eqnarray*}}
\def\eeas{\end{eqnarray*}}
\def\sla{\raise.15ex\hbox{$/$}\kern-.57em}

\def\parp{\partial^+}

\usepackage{amsmath}
\usepackage{latexsym}

\begin{document}
\begin{titlepage}
\begin{flushright}    UFIFT-HEP-05-01 \\ 
\end{flushright}
\vskip 1cm
\centerline{\LARGE{\bf {Eleven-Dimensional Supergravity in}}}
\vskip .5cm
\centerline{\LARGE{\bf {Light-Cone Superspace}}}

\vskip 1.5cm
\centerline{\bf Sudarshan Ananth\footnote{Supported by a McLaughlin Fellowship from the University of Florida and in part by the US Department of Energy under grant DE-FG02-97ER41029}, } 
\vskip .5cm
\centerline{\em  Institute for Fundamental Theory,}
\centerline{\em Department of Physics, University of Florida}
\centerline{\em Gainesville FL 32611, USA}
\vskip .5cm
\centerline{\bf Lars Brink,  }
\vskip .5cm
\centerline{\em Department of Theoretical Physics}
\centerline{\em Chalmers University
of Technology, }
\centerline{\em S-412 96 G\"oteborg, Sweden}

\vskip .5cm
\centerline{\bf Pierre Ramond${}^{\,}$\footnote{Supported in part
by the US Department of Energy under grant DE-FG02-97ER41029} }
\vskip .5cm
\centerline{\em  Institute for Fundamental Theory,}
\centerline{\em Department of Physics, University of Florida}
\centerline{\em Gainesville FL 32611, USA}

\vskip 1.5cm

\centerline{\bf {Abstract}}
\vskip .5cm
\noindent We show that Supergravity in eleven dimensions can be described in terms of a constrained superfield on the light-cone, {\it {without the use of auxiliary fields}}. We build its action to first order in the gravitational coupling constant $\kappa$, by ``oxidizing" $({\mathcal N}=8\ ,d=4)$ Supergravity. This is simply achieved, as for ${\mathcal N}=4$ Yang-Mills, by extending the transverse derivatives into superspace. The eleven-dimensional SuperPoincar\'e algebra is constructed and a fourth order interaction is conjectured.

\end{titlepage}

\section{Introduction}

${\mathcal N}=1$ supergravity in eleven dimensions~\cite{Julia}, is the largest supersymmetric local field theory with maximum helicity two (on reduction to $d=4$). This theory has gained renewed prominence since its recognition as the infrared limit of M-Theory~\cite{WITTEN}. Although M-Theory casts well-defined shadows on lower-dimensional manifolds, its actual structure remains a mystery. We must therefore glean all we can from the ${\mathcal N}=1$ supergravity theory before tackling M-Theory. ${\mathcal N}=1$ supergravity is ultraviolet divergent in $d=11$ but this divergence is presumably tamed by M-Theory and the hope is that an understanding of this divergent structure, will give us a window into the workings of M-Theory.    

Since M-Theory resides in $d=11$, we expect special features of eleven-dimensional spacetime to be reflected in its physical little group, $SO(9)$. Curtright~\cite{CURTRIGHT} conjectured that the divergences that occur in $({\mathcal N}=8,\,d=4)$ supergravity (obtained by reducing $d=11$ supergravity) were a direct consequence of the group-theoretical properties of $SO(9)$. Indeed, the mismatch between the eighth-order bosonic and fermionic Dynkin indices seems to support this conjecture. The best approach to divergence-analysis, keeping the role of the little group apparent, is on the light-cone. We therefore need to formulate eleven-dimensional supergravity in light-cone gauge before starting an analysis of its divergences. Gravity in light-cone gauge has been studied previously by numerous authors (a partial list includes,~\cite{SSCHWARZ},~\cite{KAKU},~\cite{GSCHWARZ},~\cite{ARAG}).

This paper represents the second step in our research program, initiated in reference \cite{ABR}, where we described the oxidation of $({\mathcal N}=4,d=4)$ SuperYang-Mills into $({\mathcal N}=1,d=10)$ Yang-Mills. In this paper, we start by building the eleven-dimensional SuperPoincar\'e generators. We then generalize the notion of the transverse derivative into superspace and show that this simple generalization, uniquely oxidizes $({\mathcal N}=8,d=4)$ Supergravity into the fully eleven-dimensional ${\mathcal N}=1$ Supergravity (up to first order in the coupling constant). 

In a recent paper Metsaev~\cite{MET} discussed the same theory in a somewhat different light-cone formulation, which was originally used by Brink, Green and Schwarz~\cite{Green:1983hw} in their formulation of a light-cone gauge field theory for the $d=10$ superstring. We believe that the two approaches are valuable complements in the study of $d=11$ supergravity.

\newpage

\section{The $LC_2$ Formulation of ${\mathcal N}=8$}
Our starting point for this paper is the 1982 light-cone formulation initiated in reference~\cite{CHALMERS1}. Based on the fact that both the $({\mathcal N}=8,\,d=4)$ Supergravity theory and the $({\mathcal N}=4,\,d=4)$ SuperYang-Mills theory are maximally supersymmetric, these authors introduced a constrained superfield for the ${\mathcal N}=8$ theory. We start with this superfield and show that it may be used to describe the $({\mathcal N}=1\,,d=11)$ Supergravity theory in light-cone superspace.

\subsection{Field Content}
${\mathcal N}=1$ Supergravity in eleven dimensions, contains three different massless fields, two bosonic (gravity and a three-form) and one Rarita-Schwinger spinor. Its physical degrees of freedom are classified in terms of the transverse little group, $SO(9)$, with the Graviton $G^{(\,MN\,)}$, transforming as a symmetric second-rank tensor, the three-form $B^{[\,MNP\,]}$ as an anti-symmetric third-rank tensor and the Rarita-Schwinger field as a spinor-vector, $\Psi^M$($M,N,\ldots$ are $SO(9)$ indices). This theory on reduction to four dimensions leads to the maximally supersymmetric ${\mathcal N}=8$ theory.

\noindent In $d\,=\,4$, any massless particle can be described by a  complex  field, and its complex conjugate of opposite helicity under the $SO(2)$ little group. In this case, the $SO(2)$ comes from the decomposition,
\be
SO(9)\supset~SO(2)~\times~SO(7)\ .
\ee
All told, after decomposition, the ${\mathcal N}=8$ theory has a spectrum comprised of a metric, twenty-eight vector fields, seventy scalar fields, fifty-six spin one-half fields, eight spin three-half fields and their conjugates~\cite{VANN}. The $SO(7)$ symmetry is an internal one and can in fact be upgraded to an $SU(8)$ symmetry. However, it is important to remember that it is really the $SO(7)$ which is relevant when we ``oxidize" the theory to $d\,=\,11$.

{\it All} the physical degrees of freedom of the ${\mathcal N}=8$ theory are captured by a single complex superfield~\cite{CHALMERS1},

\bea
\phi\,(y)\,&=&\,\frac{1}{{\parp}^2}\,h\,(y)\,+\,i\,{\theta^{\alpha}}\,{\frac {1}{{\parp}^2}}\,{{\bar \psi}_{\alpha}}\,(y)\,+\,i\,{\theta^{\alpha\,\beta}}\,{\frac {1}{\parp}}\,{{\bar A}_{\alpha\,\beta}}\,(y)\nonumber \\
&&-\,{\theta^{\alpha\,\beta\,\gamma}}\,{\frac {1}{\parp}}\,{{\bar \chi}_{\alpha\,\beta\,\gamma}}\,(y)-\,{\theta^{\alpha\,\beta\,\gamma\,\delta}}\,{C_{\alpha\,\beta\,\gamma\,\delta}}\,(y)\,+\,i\,{\tilde \theta}^{(5)}_{\alpha\,\beta\,\gamma}\,{\chi^{\alpha\,\beta\,\gamma}}(y) \nonumber \\
&&+\,i\,{\tilde \theta}^{(6)}_{\alpha\,\beta}\,\parp\,A^{\alpha\,\beta}(y)+\,{\tilde \theta}^{(7)}_{\alpha}\,\parp\,\chi^{\alpha}(y)\,+\,{\tilde \theta}^{(8)}\,{\parp}^2\,{\bar h}\,(y)\ ,
\eea
where,
\bea
\begin{split}
\theta^{{\alpha_1}\,\ldots\,{\alpha_n}}\,&\equiv\,\frac{1}{n!}\,{\theta^{\alpha_1}}\,\ldots\,{\theta^{\alpha_n}}\ ,\;\;\;\;\;
{\tilde {\theta}}^{(n)}_{{\beta_1}\,\ldots\,{\beta_{8-n}}}\,&\equiv\,\frac{1}{n!}\,\theta^{{\alpha_1}\,\ldots\,{\alpha_n}}\,\epsilon_{\alpha_1\,\ldots\,\alpha_n\,\beta_1\,\ldots\,\beta_{8-n}}\ .
\end{split}
\eea
 The $\gamma$-matrices allow us to translate bi-spinor indices into vector indices and vice versa (For our notation and light-cone conventions, we refer the reader to the Appendix). For example, the complex ${\bar A}_{\alpha\,\beta}$ represent,
\bea
\theta^{\alpha\,\beta}\,{{\bar A}_{\alpha\,\beta}}\,=\,-\,\frac{1}{8}\,\theta\,\gamma^m\,\theta\;{\bar A}_m\,-\,\frac{1}{8}\,\theta\,\gamma^{mn}\,\theta\;{\bar A}_{[mn]},
\eea
while the real $C_{\alpha\,\beta\,\gamma\,\delta}$ may be decomposed as,
\bea
\begin{split}
\theta^{\alpha\,\beta\,\gamma\,\delta}\,C_{\alpha\,\beta\,\gamma\,\delta}\,=\,\frac{1}{64}\,\theta\,\gamma^m\,\theta\;{\biggl \{}\,&\theta\,\gamma^p\,\theta\,C_{(mp)}\,-\,\theta\,\gamma^{mp}\,\theta\,C_p\,- \\
&\,-\,\,\theta\,\gamma^{pq}\,\theta\,C_{[mpq]}\,{\biggr \}}.
\end{split}
\eea

\noindent These tensor fields, make up the two bosonic representations under the decomposition $SO(9)\,{\subset}\,SO(7)\,\times\,SO(2)$ of eleven-dimensional supergravity. The $G^{(MN)}$ (${\bf {44}}$ of $SO(9)$), split up as,
\bea
{\bf {44\;=\;27\,+\,7+7\,+\,1+1\,+\,1}}.
\eea
$C_{(mp)}$ represents the ${\bf {27}}$ while ${\bar A}_m$, ${A^m}$ represent the ${\bf {7}}+{\bf {7}}$. Similarly, the three-form $B^{[\,MNP\,]}$ (${\bf {84}}$ of $SO(9)$) splits into,
\bea
{\bf {84\;=\;35\,+\,21+21\,+\,7}}.
\eea
These correspond to $C_{[mpq]}$, ${\bar A}_{[mn]}$, $A_{[mn]}$ and $C_p$ respectively.
\noindent All fields are local in the  modified light-cone coordinates  

\bea
y~=~\,(\,x,\,{\bar x},\,{x^+},\,y^-_{}\equiv {x^-}-\,\frac{i}{\sqrt 2}\,{\theta_{}^\alpha}\,{{\bar \theta}^{}_\alpha}\,)\ .
\eea
In this $LC_2$ form, all the unphysical degrees of freedom have been integrated out. The superfield $\phi$ and its complex conjugate $\bar\phi$ satisfy the chiral constraints,

\be
{d^{\alpha}}\,\phi\,=\,0\ ;\qquad {\bar d_{\alpha}}\,\bar\phi\,=\,0\ ,
\ee
and are related through the ``inside-out" constraints,

\be
\bar d_{\alpha}^{}\,\bar d_{\beta}^{}\,\bar d_{\gamma}^{}\,\bar d_{\eta}^{}\,\phi~=~\frac{1}{ 2}\,\epsilon_{{\alpha}{\beta}{\gamma}{\eta}{\rho}{\sigma}{\xi}{\chi}}^{}\,d^{\rho}_{}\,d^{\sigma}_{}\,d^{\xi}_{}\,d^{\chi}_{}\,\bar\phi\ ,
\ee

\noindent The ${\mathcal N}=8$ Supergravity action, to order $\kappa$ is then simply~\cite{BBB},

\be\int d^4x\int d^8\theta\,d^8 \bar \theta\,{\cal L}\;\equiv\;\int\,{\cal L}\;,
\ee
where,
\bea
\label{one}
{\cal L}&=&-\bar\phi\,\frac{\Box}{\partial^{+4}}\,\phi+(\,\frac{4\,\kappa}{3\,{{\parp}^4}}\,{\nbar \phi}\,{\bar \partial}{\bar \partial}\,{\phi}\,{{\parp}^2}{\phi}-\frac{4\,\kappa}{3\,{{\parp}^4}}\,{\nbar \phi}\,{\bar \partial}\,{\parp}{\phi}\,{\bar \partial}\,{\parp}\,{\phi}\,+\,c.c.\,)
\eea
Grassmann integration is normalized so that $\int d^8\theta\,{(\theta)}^8=1$.
\vskip 0.5cm
\noindent{\bf {A Simpler ${\mathcal N}=8$ Three-Point Vertex}}
\vskip 0.2cm
The three-point vertex in this action, seems highly non-local and cumbersome. However, its form can be greatly simplified, leading to a single term, very similar to that in the Yang-Mills case~\cite{ABR}. We start by partially integrating the first term with respect to $\bar \partial$, to obtain
\bea
\label{two}
\int\;{\biggl \{}\;-\,{\frac {\bar \partial}{{\parp}^4}}\;{\nbar \phi}\;\;{\bar \partial}\,{\phi}\;{{\parp}^2}\,{\phi}-{\frac {1}{{\parp}^4}}\;{\nbar \phi}\;\;{\bar \partial}\,{\phi}\;{\bar \partial}\,{{\parp}^2}\,{\phi}\;{\biggr \}}.
\eea
The last $\phi$ in the first term of equation (\ref{two}), may be rewritten as a $\nbar \phi$, using the inside-out relation, 
\bea
\,{\phi}\,=\,\frac{1}{2\,\cdot\,8!}\,{(d)}^8\,{\frac {1}{{\parp}^4}}{\bar \phi}\ .
\eea
We then partially integrate the ${(d)}^8$ onto the $\nbar \phi$ and use the inside-out relation again. The second term in equation (\ref{two}) is partially integrated with respect to $\parp$ to yield two terms.
\vskip 0.3cm
\noindent Thus the first term in the 3-point vertex is now,
\bea
\int\;{\biggl \{}\;-\,{\bar \partial}\;{\phi}\;\;{\bar \partial}\,{\phi}\;{\frac {1}{{\parp}^2}}\,{\nbar \phi}+{\frac {1}{{\parp}^3}}\;{\nbar \phi}\;\;{\bar \partial}\,{\phi}\;{\bar \partial}\,{\parp}\,{\phi}+{\frac {1}{{\parp}^4}}\;{\nbar \phi}\;\;{\bar \partial}\,{\parp}\,{\phi}\;{\bar \partial}\,{\parp}\,{\phi}\;{\biggr \}}.
\eea
The third term in this equation exactly cancels the second term in the original vertex. Next, we eliminate the middle term in the equation above, by recognizing that,
\bea
I\;=\;\int\,{\frac {1}{{\parp}^3}}\;{\nbar \phi}\;\;{\bar \partial}\,{\phi}\;{\bar \partial}\,{\parp}\,{\phi}\;=\;-\,\int\,{\frac {1}{{\parp}^2}}\;{\nbar \phi}\;\;{\bar \partial}\,{\phi}\;{\bar \partial}\,{\phi}\,-\,I,
\eea
(which follows from a partial integration of the single $\parp$ in the numerator). This allows us to set,
\bea
I\,=\,-\,\int\,{\frac {1}{2}}\,{\frac {1}{{\parp}^2}}\;{\nbar \phi}\;\;{\bar \partial}\,{\phi}\;{\bar \partial}\,{\phi}\,.
\eea
We thus obtain, a very concise three-point vertex,
\bea
\label{threeo}
-\;\frac{3}{2}\,{\kappa}\;{\int}\;{\frac {1}{{\parp}^2}}\;{\nbar \phi}\;\;{\bar \partial}\,{\phi}\;{\bar \partial}\,{\phi}+ c.c..
\eea
\noindent This simple form allows for comparison with the ${\mathcal N}=4$ SuperYang-Mills three-point vertex~\cite{CHALMERS1},
\bea
\,\frac {4}{3}\,g\;{\int}\;f^{abc}\,\frac {1}{\parp}\;{\nbar \phi}^a\;\phi^b\;{\bar \partial}\,\phi^c+ c.c..
\eea
We will exploit this similarity in structure, between the ${\mathcal N}=8$ and ${\mathcal N}=4$ cases, when tackling the four-point vertex.
\vskip 0.5cm

\subsection{SuperPoincar\'e Algebra}
As usual we will consider the generators of the SuperPoincar\'e algebra at the light-cone time $x^+=0$. The SuperPoincar\'e algebra, splits up into kinematical and dynamical pieces. The kinematical generators are, \newline
$\\ \bullet$ the three momenta,
\be
p^+_{}~=~-i\,\partial^+_{}\ ,\qquad p~=~-i\,\partial\ ,\qquad \bar p~=~-i\,\bar\partial\ ,
\ee
\newline
$\bullet$ the transverse space rotation,
\be
j~=~x\,\bar\partial-\bar x\,\partial+ S^{12}_{}\ ,
\ee
where, 
\be
S^{12}_{}~=~ \,\frac{1}{ 2}\,(\,{\theta^\alpha}\,{{\bar \partial}_\alpha}\,-\,{{\bar \theta}_\alpha}\,{\partial^\alpha}\,)\,+\frac{i}{4\sqrt{2}\,\partial^+}\,(\,d^\alpha\,\bar d_\alpha-\bar d_\alpha\,d^\alpha\,)\ .
\ee
This form is different from that in reference~\cite{BBB} through the last term, which acts as a helicity counter. It also ensures that the chirality constraints are preserved,
\be
[\,j\,,\,d^\alpha_{}\,]~=~[\,j\,,\,\bar d^{}_\beta\,]~=~0\ .
\ee
\newline
$\bullet$ and the ``plus-rotations",
\be
j^+_{}~=~i\, x\,\partial^+_{}\ ,\qquad \bar j^+_{}~=~i\,\bar x\,\partial^+_{}\ .
\ee
\be
 j^{+-}_{}~=~i\,x^-_{}\,\partial^+_{}-\frac{i}{2}\,(\,\theta^\alpha_{}\bar\partial^{}_\alpha+\bar\theta^{}_\alpha\,\partial^\alpha_{}\,)\ ,
\ee
which satisfy,
\bea
\begin{split}
[\,j^{+-}_{}\,,\,y^-_{}\,\,]~&=~-i\,y^-_{}\ , \\
{[\,j^{+-}_{}\,,\,d^\alpha_{}\,\,]}~&=~\frac{i}{2}\,d^\alpha_{}\ ,\qquad {[\,j^{+-}_{}\,,\,\bar d_\beta^{}\,]}~&=~\frac{i}{2}\,\bar d^{}_\beta\ ,
\end{split}
\eea
and thus preserve chirality. Note that it is only for the choice $x^+=0$ that the generator $ j^{+-}$ is kinematical, since the dynamical part is multiplied by $x^+=0$.
\vskip 0.3cm
\noindent The dynamical generators are, \newline
$\\ \bullet$ the light-cone Hamiltonian,
\be
p^-_{}~=~-i\frac{\partial\bar\partial}{\partial^+_{}}
\ee
\newline
$\bullet$ and the dynamical boosts,
\bea
j^-_{}&=&i\,x\,\frac{\partial\bar\partial}{\partial^+_{}} ~-~i\,x^-_{}\,\partial~+~i\,\Big( \theta^\alpha_{}\bar\partial^{}_\alpha\,+\frac{i}{4\sqrt{2}\,\partial^+}\,(\,d^\alpha\,\bar d_\alpha-\bar d_\alpha\,d^\alpha\,)\Big)\frac{\partial}{\partial^+_{}}\,\ ,\cr 
\bar j^-_{}&=&i\,\bar x\,\frac{\partial\bar\partial}{\partial^+_{}}~ -~i\,x^-_{}\,\bar\partial~+~ i\,\Big(\bar\theta_\beta^{}\partial_{}^\beta+\frac{i}{4\sqrt{2}\,\partial^+}\,(\,d^\beta\,\bar d_\beta-\bar d_\beta\,d^\beta\,)\,\Big)\frac{\bar\partial}{\partial^+_{}}\,\ .
\eea
The helicity counter added to the expressions above, once again ensures that they leave chirality and the inside-out relations unaltered because, 
\be
[\,j^{-}_{}\,,\,d^\alpha_{}\,]~=~\frac{i}{2}\,d^\alpha_{}\,\frac{\partial}{\partial^+_{}}\ ,\qquad [\,j^{-}_{}\,,\,\bar d_\beta^{}\,]~=~\frac{i}{2}\,\bar d_\beta^{}\,\frac{\partial}{\partial^+_{}}\ .
\ee
These generators also satisfy,

\be
[\,j_{}^-\,,\,\bar j^+_{}\,]~=~-i\,j^{+-}_{}-j\ ,\qquad [\,j^-_{}\,,\,j^{+-}_{}\,]~=~i\,j^{-}_{}\ .
\ee
\newline
In a similar fashion, the supersymmetries split into,\newline
$\\ \bullet$ kinematical supersymmetries,

\be
q^{\,\alpha}_{\,+}=-{\partial^\alpha}\,+\,\frac{i}{\sqrt 2}\,{\theta^\alpha}\,{\partial^+}\ ;\qquad{{\bar q}_{\,+\,\beta}}=\;\;\;{{\bar \partial}_\beta}\,-\,\frac{i}{\sqrt 2}\,{{\bar \theta}_\beta}\,{\partial^+}\ ,
\ee
satisfying
\be
\{\,q^{\,\alpha}_{\,+}\,,\,{{\bar q}_{\,+\,\beta}}\,\}\,=\,i\,{\sqrt 2}\,{{\delta^\alpha}_\beta}\,{\parp}\ ,
\ee
and anticommuting with the chiral derivatives
\bea
\{\,q^{\,\alpha}_{\,+}\,,\,{{\bar d}_\beta}\,\}\,=\,\{\,{d^\alpha}\,,\,{{\bar q}_{\,+\,\beta}}\,\}\,=\,0\ .
\eea
\newline
$\bullet$ dynamical supersymmetries, which may be obtained by boosting the kinematical ones
\be
\label{dynfs}
{q}^\alpha_{\,-}~\equiv~i\,[\,\bar j^-_{}\,,\,q^{\,\alpha}_{\,+}\,]~=~\frac{\bar \partial}{\partial^+_{}}\, q^{\,\alpha}_{\,+}\ ,\qquad 
{\bar{q}}_{\,-\,\beta}^{}~\equiv~i\,[\, j^-_{}\,,\,\bar q_{\,+\,\beta}^{}\,]~=~\frac{\partial}{\partial^+_{}}\, \bar q_{\,+\,\beta}^{}\ .
 \ee
They obey,
\be
\{\,{q}^\alpha_{\,-}\,,\,{\bar{q}}_{\,-\,\beta}^{}\,\}~=~i\,\sqrt{2}\,\delta^{\,\alpha}_{~~\beta}\,\frac{\partial\bar\partial}{\partial^+_{}}\ ,
\ee 
\be
\{{q}^\alpha_{\,+}\,,\,{\bar{q}}_{\,-\,\beta}^{}\,\}~=~i\,\sqrt{2}\,\delta^{\,\alpha}_{~~\beta}\,\partial\ .
\ee
Except for the changes to ensure chirality, these operators all appear in reference~\cite{BBB}. In the next section, we generalize this four-dimensional algebra to eleven dimensions.

\section{Eleven Dimensions}
The main aim of this paper is to show that the fully interacting $({\mathcal N}=8,d=4)$ theory can be restored to its eleven-dimensional progenitor, without altering the superfield. This enables us to formulate the $({\mathcal N}=1,d=11)$ theory without auxiliary fields. 

The first step is to generalize the transverse variables to nine. In the Yang-Mills case, the compactified $SO(6)$ was easily described by $SU(4)$ parameters and we made use of the convenient bi-spinor notation. In the present case, the compactified $SO(7)$ has no equivalent unitary group so we simply introduce additional real coordinates, ${x^m}$ and their derivatives $\partial^m$(where $m$ runs from $4$ through $10$). The chiral superfield remains unaltered, except for the added dependence on the extra coordinates
\be
h(y)~=~h(x,\bar x,{x^m},y^-_{})\ ,~~etc... ~\ .
\ee 
These extra variables  will be acted on by new operators that will restore the higher-dimensional symmetries.

\subsection{The SuperPoincar\'e Algebra in $11$ Dimensions}
The SuperPoincar\'e algebra needs to be generalized from its four-dimensional version. The $SO(2)$ generators stay the same and we propose generators of the coset $SO(9)/(SO(2)\times SO(7))$, of the form,
\bea
{J^m}\,&=&\,-\,i\,(\,x\,{\partial^m}\,-\,{x^m}\,{\partial}\,)\,+\,{\frac {i}{2\,\sqrt 2}}\,\parp\;{\theta^\alpha}\,{{(\,{\gamma^m})}_{\alpha\,\beta}}\,{\theta^\beta}\,-\,{\frac {i}{\sqrt 2\,\parp}}\;{\partial^\alpha}\,{{(\,{\gamma^m})}_{\alpha\,\beta}}\,{\partial^\beta}\, \nonumber \\
&&+\,{\frac {i}{2\,\sqrt 2\,{\parp}}}\;{d^\alpha}\,{{(\,{\gamma^m})}_{\alpha\,\beta}}\,{d^\beta}
\eea

\bea
{{\nbar J}^{\;n}}\,&=&\,-\,i\,(\,{\bar x}\,{\partial^n}\,-\,{x^n}\,{\bar \partial}\,)\,+\,{\frac {i}{2\,\sqrt 2}}\,\parp\;{{\bar \theta}_\alpha}\,{{(\,{\gamma^n})}^{\alpha\,\beta}}\,{{\bar \theta}_\beta}\,-\,{\frac {i}{\sqrt 2\,\parp}}\;{{\bar \partial}_\alpha}\,{{(\,{\gamma^n})}^{\alpha\,\beta}}\,{{\bar \partial}_\beta}\, \nonumber \\
&&+\,{\frac {i}{2\,\sqrt 2\,{\parp}}}\;{{\bar d}_\alpha}\,{{(\,{\gamma^n})}^{\alpha\,\beta}}\,{{\bar d}_\beta}
\eea

\noindent which satisfy the $SO(9)$ commutation relations,

\bea
\Big[\,J\,,\,J^m\,\Big]&=&J^m\ ,\qquad \Big[\,J\,,\,\bar J^n\,\Big]~=~-\bar J^n \nonumber \\
\Big[\,J^{pq}\,,\,J^m\,\Big]&=&\delta^{pm}\,J^q\,-\,\delta^{qm}\,J^p \nonumber \\
{\Big[}\,J^m\,,\,\bar J^n\,{\Big ]}&=&\,i\,J^{mn}\,+\,\delta^{mn}\,J,
\eea
where $J$ is the same as before, $J\,=\,j$ and the $SO(7)$ generators read,
\bea
J^{mn}\,&=&\,-\,i\,(\,x^m\,\partial^n\,-\,x^n\,\partial^n\,)\,+\,\theta^\alpha\,{(\gamma^m)}^{\alpha\,\,\beta}\,{(\gamma^n)}^{\beta\,\,\sigma}\,{\bar \partial}_\sigma\, \nonumber \\
&&+\,{\bar \theta}_\alpha\,{(\gamma^m)}^{\alpha\,\,\beta}\,{(\gamma^n)}^{\beta\,\,\sigma}\,{\partial}^\sigma\,-\frac{1}{\sqrt 2\,\parp}\,{d^\alpha}\,{(\gamma^m)}^{\alpha\,\,\beta}\,{(\gamma^n)}^{\beta\,\,\sigma}\,{\bar d}_\sigma\ .
\eea
\noindent The full $SO(9)$ transverse algebra is generated by $J\,,\,J^{mn}\,,\,{J^m}$ and ${\bar J}^n$. All rotations are specially constructed to preserve chirality. For example,
\be
\,[\,{J^{m}}\,,\,{{\bar d}_{\alpha}}\,]\,~=~0\ ;\qquad \,[\,{{\bar J}^{n}}\,,\,{d^\alpha}\,]\,~=~0\ .
\ee
The remaining kinematical generators do not get modified,
\bea
J^+\,=\,j^+\ ,\qquad J^{+-}\,=\,j^{+-}\ ,
\eea
while new kinematical generators appear,
\bea
J^{+\,m}&=&i\,x^{m}\,\partial^+_{}\ ; \qquad 
\bar J^{+\,n}~=~i\,\bar x^{n}\,\partial^+_{}\ . 
\eea
We generalize the linear part of the dynamical boosts to, 
\bea
J^-_{}&=&i\,x\,\frac{\partial\bar\partial\,+\,{\frac {1}{2}}\,{\partial^m}\,{\partial^m}}{\partial^+_{}} ~-~i\,x^-_{}\,\partial+i\,{\frac { \partial}{\parp}}\,\Big\{\,{ \theta}^\alpha\,{\bar\partial_\alpha}~+~\frac{i}{4\sqrt{2}\,\parp}\,(d^\alpha\,\bar d_\alpha-\bar d_\alpha\,d^\alpha)\,\Big\} \nonumber \\
&&-\,{\frac {1}{4}}\,{\frac {\partial_m}{\parp}}\,{\biggl \{}\,{\parp}\;{\theta^\alpha}\,{{(\,{\gamma^m})}_{\alpha\,\beta}}\,{\theta^\beta}\,-\,{\frac {2}{\parp}}\,\;{\partial^\alpha}\,{{(\,{\gamma^m})}_{\alpha\,\beta}}\,{\partial^\beta}\,+\,{\frac {1}{\parp}}\,\;{d^\alpha}\,{{(\,{\gamma^m})}_{\alpha\,\beta}}\,{d^\beta}\,{\biggr \}} \nonumber \\
\,
\eea

\noindent The other boosts may be obtained by using the $SO(9)/(SO(2)\times SO(7))$ rotations,

\be
J^{-\,m}_{}~=~[\,J^-_{}\,,\,J^{m}_{}\,]\ ;\qquad 
\bar J^{-\,n}~=~[\,\bar J^-_{}\,,\,\bar J^{n}\,]\ .
\ee
We do not show their explicit forms as they are too cumbersome. The dynamical supersymmetries are obtained by boosting

\bea
\label{ds}
[\,J^-\,,\,\bar q_{+\,\eta}\,]~\equiv~\nbar{\cal Q}_\eta^{}&=&-\,i\,\frac {\partial}{\parp}}\,{{\nbar q}_{+\,\eta}}\,-\,{\frac{i}{\sqrt 2}}\,{{(\,{\gamma^n}\,)}_{\,\eta\,\rho}}\,q^{~\rho}_{\,+}\,{\frac {\partial^n}{\parp}\ ,\cr
&&\cr
[\,{\bar J}^-\,,\,q_+^\alpha\,]~\equiv~{\cal Q}^\alpha_{}&=&i\,{\frac {\bar \partial}{\parp}}\,{{q_+}^\alpha}\,+\,{\frac {i}{\sqrt 2}}\,{{(\,{\gamma^m}\,)}^{\,\alpha\,\beta}}\,{{\bar q}_{+\,\beta}^{}}\,{\frac {\partial^m}{\parp}}\ .
\eea

\noindent They satisfy, 
\bea
\{\,{\cal Q}^{\,\alpha}_{}\,,\,q_+^\eta\,\}~=~-\,{{(\,{\gamma^m}\,)}^{\alpha\,\eta}}\,{\partial^m}\ ,
\eea

\noindent and the supersymmetry algebra,

\bea
\{\,{\cal Q}^{\,\alpha}_{}\,,\,\nbar {\cal Q}^{}_{\,\eta}\,\}~=~i\,\sqrt{2}\,\;{{\delta^{\alpha}}_{\eta}}\,\frac{1}{\parp}
\,\Big(\partial\,{\nbar \partial}\,+\,\frac{1}{2}\,{\partial^m}\,{\partial^m}\,\Big)\ .
\eea
A few central charges fit nicely into this framework. The $d=11$ theory is known to have 517 central charges; $7$ of these, may be introduced by simply replacing the seven derivatives $\partial^m$, by c-numbers $Z^m$. The remaining $510$ charges have an eleven-dimensional origin. 

\noindent Having constructed the free ${\mathcal N}=1$ SuperPoincar\'e generators in eleven dimensions which act on the chiral superfield, we turn to building the interacting theory.


\subsection{The Generalized Derivatives}
The cubic interaction in the ${\mathcal N}=8$ Lagrangian explicitly contains the  transverse derivative operators $\partial$ and $\bar\partial$. To achieve covariance in eleven dimensions, we proceed to generalize these operators as we did for ${\mathcal N}=4$  Yang-Mills. We propose the generalized derivative 

\bea
{\nbar \nabla}\;=\;{\bar \partial}\,+\,{\frac {\sigma}{16}}\,{{\bar d}_{\alpha}}\,{{(\,{\gamma^m}\,)}^{\alpha\,\beta}}\,{{\bar d}_\beta}\,{\frac {\partial^m}{\parp}}\,,
\eea
which naturally incorporates the coset derivatives $\partial^{m}$. Here $\sigma$ is a parameter, still to be determined. We use the coset generators to produce its rotated partner $\nbar\nabla$ by,

\bea
[\;{\nbar \nabla}\,,\,{J^m}\;]\;{\equiv}\;{\nabla^m}\;=\;-\,i\,{\partial^m}\,+\,{\frac {i\,\sigma}{16}}\,{{\bar d}_{\alpha}}\,{{(\,{\gamma^m}\,)}^{\alpha\,\beta}}\,{{\bar d}_\beta}\,{\frac {\partial}{\parp}}\,.
\eea

\noindent It remains to verify that the original derivative operator is reproduced by undoing this rotation; indeed we find the required closure,

\beas
[\;{\nabla^m}\,,\,{{\nbar J}^{\;p}}\;]\;=\;{\delta^{\,m\,p}}\,\;{\nbar \nabla}
\eeas

\vskip 0.3cm

\noindent The new derivative $(\;{\nbar \nabla}\,,\,{\nabla^m}\;)$, thus transforms as a 9-vector under the little group in eleven dimensions. We note that $\sigma$ is not determined by these algebraic requirements. Instead, its value will be fixed in the next section, by requiring that our generalized vertex, satisfy invariance requirements. We define the conjugate derivative $\nabla$, by requiring that 

\be
\nabla\,\bar\phi~\equiv~\nbar{(\nbar\nabla\,\phi)}\ .
\ee
This tells us that, 

\bea
\nabla~\equiv~\;{\partial}\,+\,{\frac {\sigma^*}{16}}\,{d^\alpha}\,{{(\,{\gamma^n}\,)}^{\alpha\,\beta}}\,{d^\beta}\,{\frac {\partial^n}{\parp}}\,
\eea 
This construction is akin to that for the ${\mathcal N}=4$ Yang-Mills theory, but this time it applies to the ``oxidation" of the (${\mathcal N}=8$, $d=4$) theory to (${\mathcal N}=1$, $d=11$) Supergravity. This points to remarkable algebraic similarities between the two theories, with possibly profound physical consequences. It remains to show that the simple replacement of the transverse derivatives $\partial,\bar \partial$ by $\nabla,\nbar \nabla$ in the (${\mathcal N}=8$, $d=4$) interacting theory yields the fully covariant Lagrangian in eleven dimensions. 

\subsection{Invariance of the Action}
While the covariant (${\mathcal N}=8$, $d=4$) Supergravity Lagrangian is known to all orders in the gravitational coupling, its light-cone superspace expression has only been constructed to first order in $\kappa$ (three-point coupling). We note that the four-point gravity vertex is indeed known, in component form~\cite{BCL}\cite{MOG} but a superfield expression remains elusive. We will work with the theory to cubic order in this section and in the next section, propose a quartic interaction term for the theory. The kinetic term is trivially made $SO(9)$-invariant by including the seven extra transverse derivatives in the d'Alembertian. 

\noindent We start from our simplified version (\ref {threeo}) of the ${\mathcal N}=8$ three-point vertex. We propose that the eleven-dimensional vertex is of the same form, but with the transverse derivatives replaced by the generalized derivatives, introduced in the previous section: 

\bea
\begin{split}
{\mathcal V}\;=&-\;{\frac {3}{2}}\,{\kappa}\;{\int}\,{d^{11}}x\,{\int}\,{d^8}{\theta}\,{d^8}{\bar \theta}\;{\frac {1}{{\parp}^2}}\;{\nbar \phi}\;\;{\nbar \nabla}\,{\phi}\;{\nbar \nabla}\,{\phi}\ ,\\ 
\end{split}
\eea
together with its complex conjugate. To show $SO(9)$ invariance, it suffices to consider the variations

\bea
{\delta_{J^m}}\,{\phi}\;=\;\,\frac{i}{\sqrt 2}\;{\omega_m}\,{\parp}\,\;{\theta^\alpha}\,{{(\,{\gamma^m})}_{\alpha\,\beta}}\,{\theta^\beta}\,{\phi}\ ,
\eea
\bea
\begin{split}
{\delta_{J^m}}\,{\nbar \phi}\,=&\,{\omega_m}\,{\biggl \{}\,{\frac {i}{2\,\sqrt 2}}\,\parp\;{\theta^\alpha}\,{{(\,{\gamma^m})}_{\alpha\,\beta}}\,{\theta^\beta}\,-\,{\frac {i}{\sqrt 2\,\parp}}\;{\partial^\alpha}\,{{(\,{\gamma^m})}_{\alpha\,\beta}}\,{\partial^\beta}\, \\
\;\;\;\;\;\;\;\;\;&\;\;\;\;\;\;\;+\,{\frac {i}{2\,\sqrt 2\,{\parp}}}\;{d^\alpha}\,{{(\,{\gamma^m})}_{\alpha\,\beta}}\,{d^\beta}\,{\biggr \}}\,{\nbar \phi}\ ,
\end{split}
\eea
\bea
{\delta_{J^m}}\,{\nbar \nabla}\;=\;-\,{\omega_m}\,{\nabla^m}\ ,
\eea
where $\omega_m$ are the parameters of the $SO(9)/(SO(7)\times SO(2))$ coset transformations. The $SO(2)$ invariance is clear from the work in $d=4$ and the $SO(7)$ invariance is covariantly realized so if we can show the invariance under the  $SO(9)/(SO(7)\times SO(2))$ transformations we have shown invariance under the full $SO(9)$.
\subsubsection{The Variation}

We split the calculation into three parts, based on which superfield is being varied. Terms that involve ${\bar \partial}\,{\bar \partial}$ or ${\partial^m}\,{\partial^n}$, cancel trivially. 

\noindent The remaining terms all involve a single $SO(2)$ derivative and a single ${\partial^m}$. We list the contributions from the variations below,

\vskip 0.3cm

{\bf {Contribution}} from $\int\;{\frac {1}{{\parp}^2}}\;{\biggl (}\,{\delta_J}\,{\nbar \phi}\,{\biggr )}\;{\nbar \nabla}\,{\phi}\;{\nbar \nabla}\,{\phi}$:

\bea
\begin{split}
\int\;\frac{i\,\sigma}{2\,\sqrt 2}{\biggl \{}&{\frac {8}{{\parp}^3}}\,{\nbar \phi}\;{\bar \partial}\,{\phi}\,{\partial^m}\,{\parp}\,{\phi}\;-\;{\frac {4}{{\parp}^3}}\,{\nbar \phi}\;{\bar \partial}\,{\phi}\,{\partial^m}\,{\parp}\,{\phi} \\
&-\;{\frac {1}{{\parp}^3}}\,{\nbar \phi}\;{\bar \partial}\,{\phi}\,{{(\gamma^m)}^{\,\beta\kappa}}\,{{(\gamma^n)}^{\,\kappa\chi}}\,{\theta^\beta}\,{{\bar d}_\chi}\,{\partial^n}\,{\parp}\,{\phi} \\
&-\,{\frac {1}{{\parp}^3}}\,{\nbar \phi}\;{\theta^\beta}\,{\bar \partial}\,{\parp}\,{\phi}\,{{(\gamma^m)}^{\,\beta\kappa}}\,{{(\gamma^n)}^{\,\kappa\chi}}\,{{\bar d}_\chi}\,{\partial^n}\,{\phi}\;\biggr \}
\end{split}
\eea

\vskip 1cm

{\bf {Contribution}} from $\int\;{\frac {1}{{\parp}^2}}\;{\nbar \phi}\,\;{\biggl (}\,{\delta_J}\,{\nbar \nabla}\,{\biggr )}\,{\phi}\;{\nbar \nabla}\,{\phi}$:

\bea
\int\;{\frac {2\,i\,}{{\parp}^2}}\,{\nbar \phi}\;{\bar \partial}\,{\phi}\,{\partial^m}\,{\phi}
\eea

\vskip 1cm

{\bf {Contribution}} from $\int\;{\frac {1}{{\parp}^2}}\;{\nbar \phi}\,\;{\nbar \nabla}\,{\biggl (}\,{\delta_J}\,{\phi}\,{\biggr )}\;{\nbar \nabla}\,{\phi}$:

\bea
\int\;\frac{i\,\sigma}{2\,\sqrt 2}\,{\biggl \{}\;{\frac {4}{{\parp}^2}}\,{\nbar \phi}\;{\bar \partial}\,{\phi}\,{\partial^m}\,{\phi}\;-\;{\frac {1}{{\parp}^2}}\,{\nbar \phi}\;{\theta^\beta}\,{\bar \partial}\,{\phi}\,{{(\gamma^m)}^{\,\beta\kappa}}\,{{(\gamma^n)}^{\,\kappa\chi}}\,{{\bar d}_\chi}\,{\partial^n}\,{\phi}\;{\biggr \}}\ ,
\eea

\vskip 1cm
\noindent These results are further simplified, by use of the magical identity,

\bea
{\int}\,{\frac {1}{{\parp}^3}}\,{\nbar \phi}\,{\bar \partial}\,{\phi}\,{\partial^m}\,{\parp}\,{\phi}\,=\,{\int}\,{\frac {1}{{\parp}^3}}\,{\nbar \phi}\,{\partial^m}\,{\phi}\,{\bar \partial}\,{\parp}\,{\phi},
\eea
which follows from the duality constraint and numerous partial integrations.

\noindent The final form of the variation then reads,

\bea
{\delta_J}\,{\cal V}\,\propto\,\int\,{\biggl (}\;\frac{i\,\sigma}{\sqrt 2}+\,i\,{\biggr )}\;{\frac {1}{{\parp}^2}}\,{\nbar \phi}\;{\bar \partial}\,{\phi}\,{\partial^m}\,{\phi}\ ,
\eea
\noindent Eleven-dimensional Poincar\'e invariance requires that this variation vanish. 
This determines $\sigma$ and hence the generalized derivative,

\bea
{\nbar \nabla}\;=\;{\bar \partial}\,-\,{\frac {1}{8\,\sqrt 2}}\,{{\bar d}_{\alpha}}\,{{(\,{\gamma^m}\,)}^{\alpha\,\beta}}\,{{\bar d}_\beta}\,{\frac {\partial^m}{\parp}}\,.
\eea
This completes the proof of $SO(9)$ invariance for the three-point function. It is rather remarkable that this simple replacement of the transverse derivatives, renders the action covariant in eleven dimensions. 
\noindent In this light-cone form, the Lorentz invariance in eleven dimensions is automatic once the little group invariance has been established. We have therefore proven eleven-dimensional invariance to order $\kappa$.

\section{Extension to Order $\kappa^2$}

The next step in this program, is to extend the oxidation procedure to order $\kappa^2$. However, the starting point for this process, the $({\mathcal N}=8,\,d=4)$ four-point interaction (in light-cone superspace) is unknown. In this section, we outline a general procedure to derive the four-dimensional quartic interaction. Once this vertex is determined, we expect the oxidation procedure to follow along very similar lines (thus yielding, the $({\mathcal N}=1, d=11)$ action to order $\kappa^2$).

\vskip 0.7cm
\subsection{Conjectured Quartic Vertex}
\vskip 0.3cm
The similarity between the three-point functions of the ${\mathcal N}=4$ Yang-Mills and ${\mathcal N}=8$ Supergravity theories is quite suggestive. Based on this comparison, we conjecture that the $({\mathcal N}=8,d=4)$ four-point vertex is simply,
\bea
{\cal V}\;=\;{\kappa^2}\,{\int}\,\,{\biggl \{}\;{\nbar \phi}\;{\phi}\;{\bar \partial}\,\phi\,{\partial}\,{\nbar \phi}\;+\;\beta\;{\frac {1}{{\parp}^2}}\,(\,{\bar \partial}\,{\phi}\;{{\parp}^2}\,{\phi}\,)\,{\frac {1}{{\parp}^2}}\,(\,{\partial}\,{\nbar \phi}\,{{\parp}^2}\,{\nbar \phi}\,)\;{\biggr \}}
\eea
\noindent where $\beta$ remains to be fixed. A direct check of this result can be achieved by comparison with the component form of light-cone gravity. As mentioned in the previous section, the four-point function for gravity in light-cone gauge is known~\cite{BCL}, but there exists no expression for the quartic(or higher) vertices in terms of superfields in the literature. In component form, it is well known that the light-cone time derivative $\partial_+$, makes an appearance at every order and needs to be eliminated via field-redefinitions. This redefinition needs to be understood from a superspace point of view. We hope to return to these issues in a future publication~\cite{AKR}. Assuming this four-point vertex (in four dimensions), its ``oxidized" eleven-dimensional version is conjectured to be,
\bea
{\cal V}\;=\;{\kappa^2}\,{\int}\,{\biggl \{}\;{\nbar \phi}\;{\phi}\;{\nbar \nabla}\,\phi\,{\nabla}\,{\nbar \phi}\;+\;\beta\;{\frac {1}{{\parp}^2}}\,(\,{\nbar \nabla}\,{\phi}\;{{\parp}^2}\,{\phi}\,)\,{\frac {1}{{\parp}^2}}\,(\,{\nabla}\,{\nbar \phi}\,{{\parp}^2}\,{\nbar \phi}\,)\;{\biggr \}}\ .
\eea
\noindent It remains to be seen if this simple form reproduces the full quartic vertex of eleven-dimensional supergravity.

\subsection{Supersymmetry Variations}

A possible way to determine the four-point vertex is to start from the known two- and three-point vertices and require invariance of the action (to order $\kappa^2$) under the non-linear dynamical SuperPoincar\'e transformations. Simplest among these are the dynamical supersymmetry generators, already derived up to first order in the coupling constant for the $({\mathcal N}=8,d=4)$ theory in reference \cite{BBB}. 
\bea
\begin{split}
{{\delta_{{\bar q}_-}}^0}\,\phi\,&=\,\frac{\partial}{\partial^+_{}}\,{{\bar q}_+}\,\phi \\
{\delta_{{\bar q}_-}}^\kappa\,\phi\,&=\,-\,2\,\kappa\,\frac {1}{\parp}\,{\biggl \{}\,{\bar \partial}\,{\frac {\partial}{\partial\,\theta}}\,\phi\,{{\parp}^2}\,\phi\;-\;\parp\,{\frac {\partial}{\partial\,\theta}}\,\phi\,\parp\,{\bar \partial}\,\phi\,{\biggr \}}\ .
\end{split}
\eea
This variation, clearly preserves the chirality of the superfield it acts on. Note that we can replace the ${\frac {\partial}{\partial\,\theta}}$ in the above expression by $\bar d$ or ${\bar q}_+$. This is possible because the additional $\bar\theta\,\parp$ piece cancels between the two terms in the variation.
\vskip 0.3cm
\noindent To determine the four-point function, we need the dynamical supersymmetry variation at order $\kappa^2$. We do not yet know its exact form but have narrowed it down, based on the following requirements:
\begin{itemize}
\item Helicity $=\,+\,\frac{5}{2}$ (based on the variations at order $0$ and $\kappa$)
\item Dimensions $=\,{[L]}^{\frac{1}{2}}$ (again, based on the earlier variations)
\item The chirality of the superfield it acts on must not be affected
\item Finally, it must preserve the ``inside-out" superfield constraints.
\end{itemize}

The requirement that it leave chirality invariant, is satisfied through Chiralization, a procedure we explain in section {\bf {4.3}}. The first three constraints offer an Ansatz for the variation,
\bea
{\delta_{{\bar q}_-}}^{\kappa^2}\,\phi\,\propto\,{\kappa^2}\,{\frac {1}{{\parp}^n}}\,{\biggl \{}\,{\parp}^a\,{\bar d}_\beta\,\phi\,{{\parp}^b}\,\phi\,{\parp}^c\,{\partial}\,\bar\phi\,{\biggr \}}\,+\,\cdots
\eea
with the restriction,  $a\,+\,b+\,c\,=\,3\,+\,n$ (the $\cdots$ signify that more terms need to be added to ensure chirality). Once this expression is known, the four-point interaction is easy to determine. In a future publication~\cite{AKR}, we will derive the exact form of ${\delta_{{\bar q}_-}}^{\kappa^2}$ and prove in addition, closure of the supersymmetry algebra with the SuperPoincar\'e generators.
\vskip 0.3cm
\noindent The lowest order dynamical supersymmetry variations for the $({\mathcal N}=1,d=11)$ theory were detailed in equation (\ref{ds}). We conjecture that at first order in coupling, the variations are obtained by simply oxidizing the four-dimensional result above. That is,
\bea
{\delta_{{\bar {\mathcal Q}}_-}}^\kappa\,\phi&=&-\,2\,\kappa\,\frac {1}{\parp}\,{\biggl \{}\,{\nbar \nabla}\,{\frac {\partial}{\partial\,\theta}}\,\phi\,{{\parp}^2}\,\phi\;-\;\parp\,{\frac {\partial}{\partial\,\theta}}\,\phi\,\parp\,{\nbar \nabla}\,\phi\,{\biggr \}}
\eea

\noindent This result needs to be checked but we believe it to be correct since it serves as a bridge between the two- and three-point vertices.

\subsection{Chiralization}

``Chiralization" is a descent procedure whereby non-chiral expressions are rendered chiral. For any general non-chiral expression of the form, $A\,{\bar B}$ (where $A$ is any compound chiral function and $\bar B$ a compound anti-chiral function), we define a ``chiral product" through a descent relation in chiral derivatives,
\bea
{\cal C}\,(\,A{\bar B}\,)\,=\,
A\,{\bar B}\,+\,{\sum_{n=1}^8}\,{\frac {{(-1)}^n}{n!}}\,{\frac {{\bar d}_{\alpha_1\,\ldots\,\alpha_n}}{{(-\,i\,{\sqrt 2}\,\parp)}^n}}\;(\,A\,{d^{\alpha_n\,\ldots\,\alpha_1}}\,{\bar B}\,),
\eea
where ${\bar d}_{\alpha_1\,\ldots\,\alpha_n}\,=\,{{\bar d}_{\alpha_1}}\,\ldots\,{{\bar d}_{\alpha_n}}$ and ${d^{\alpha_n\,\ldots\,\alpha_1}}\,=\,d^{\alpha_n}\,\ldots\,d^{\alpha_1}$.

\vskip 0.3cm
\noindent ${\cal C}\,(\,A{\bar B}\,)\,$ is now a chiral function and satisfies, $d\,{\cal C}\,=\,0$. Clearly, the descent series involves as many terms as there are supersymmetries in the theory.
\vskip .3cm
\noindent This procedure is equally applicable to other non-chiral forms. For example, the product, ${\bar d}\,A\,B$ where both $A$ and $B$ are chiral functions, is chiralized by the addition of the term,
\bea
-\,\parp\,A\,\frac{\bar d}{\parp}\,B
\eea
\vskip 0.5cm
\noindent Similarly, the addition of the two terms,
\bea
-\,\parp\,\bar d\,A\,\frac{\bar d}{\parp}\,B\;+\;{\parp}^2\,A\,\frac{\bar d\,\bar d}{{\parp}^2}\,B
\eea
to the expression $\bar d\,\bar d\,A\,B$, chiralizes it and so on.
\vskip 0.5cm
This procedure is invaluable when dealing with variations (with respect to supersymmetry and the boosts) at higher orders. This simple recipe ensures that all variations, respect the superfield chirality structure.

\section{Conclusions}
Eleven-dimensional Supergravity has been successfully formulated in light-cone superspace to order $\kappa$. Once again, the key to this formulation is the generalized derivative whose components transform as an $SO(9)$ vector. Further progress necessarily requires the determination of the quartic interaction. We will describe the procedure to derive these higher-point vertices in superspace, in a future publication~\cite{AKR}.
\vskip 0.3cm
Based on Curtright's reasoning\cite{CURTRIGHT}, the divergence of the $({\mathcal N}=8,d=4)$ theory may be attributed to the incomplete cancelation of the Dynkin indices(starting with the eighth-order invariant) of the $SO(9)$ representations. The eighth-order invariant first appears at three loops and hence the divergent nature of the theory necessarily begins at three loops or higher. Our formulation of the $\mathcal N=8$ theory provides an ideal framework to verify this explicitly, because the role of the little-group is made apparent on the light-cone. We hope to come back to this analysis and compare our results with existing predictions in the literature. In reference~\cite{DIXON} for example, the authors argue that no three-loop divergence appears and further, that the first counterterm occurs at five loops.
{\vskip 1cm}

\noindent {\bf {Acknowledgments}}\\[0.5cm]
We would like to thank Sung-Soo Kim for many helpful discussions and reading the manuscript. 

\vskip 1cm

\renewcommand{\theequation}{A-\arabic{equation}}
  \setcounter{equation}{0}  
  \section*{APPENDIX}  

With the space-time metric $(-,+,+,\dots,+)$, the light-cone coordinates and their derivatives are 
\bea
{x^{\pm}}&=&\frac{1}{\sqrt 2}\,(\,{x^0}\,{\pm}\,{x^3}\,)\ ;\qquad ~ {\partial^{\pm}}=\frac{1}{\sqrt 2}\,(\,-\,{\partial_0}\,{\pm}\,{\partial_3}\,)\ ; \\
x &=&\frac{1}{\sqrt 2}\,(\,{x_1}\,+\,i\,{x_2}\,)\ ;\qquad  {\bar\partial} =\frac{1}{\sqrt 2}\,(\,{\partial_1}\,-\,i\,{\partial_2}\,)\ ; \\
{\bar x}& =&\frac{1}{\sqrt 2}\,(\,{x_1}\,-\,i\,{x_2}\,)\ ;\qquad  {\partial} =\frac{1}{\sqrt 2}\,(\,{\partial_1}\,+\,i\,{\partial_2}\,)\ ,
\eea
such that 
\be
{\parp}\,{x^-}={\partial^-}\,{x^+}\,=\,-\,1\ ;\qquad {\bar \partial}\,x\,=\,{\partial}\,{\bar x}\,=+1 \ .
\ee

Anti-commuting Grassmann variables $\theta^{\alpha}$ and their conjugates, $\bar\theta_\alpha$($\alpha\,=\,1,2,\ldots,8$) are defined by, 
\be
\{\,{\theta^{\alpha}}\,,\,{{ \theta}^\beta}\,\}\,~=~\,\{{{\bar \theta}_\alpha}\,,\,{\bar\theta_\beta}\,\}\,~=~\,\{{{\bar \theta}_\alpha}\,,\,{\theta^\beta}\,\}\,~=~0\ ,
\ee
$\theta^\alpha$ transforms according to the real spinor representation of $SO(7)$. Despite the reality of the spinor representation, we continue to use the ``up" and ``down" indices for notational convenience. Their derivatives are written as 

\be
{{\bar \partial}_{\alpha}}\,~\equiv~\,\frac{\partial}{\partial\,{\theta^{\alpha}}}\ ;\qquad{\partial^{\beta}}\,~\equiv~\,\frac{\partial}{\partial\,{\bar \theta}_{\beta}}\ ,
\ee
with canonical anticommutation relations

\be
\{\,{\partial^{\alpha}}\,,\,{{\bar \theta}_{\beta}}\,\}\,~=~\,{{\delta^{\alpha}}_\beta}\ ;\qquad \{\,{{\bar \partial}_{\alpha}}\,,\,{\theta^{\beta}}\,\}\,~=~\,{{\delta_{\alpha}}^{\beta}}\ .
\ee
Under conjugation, upper and lower spinor indices are interchanged, so that $\nbar{{\theta^{\alpha}}}\,=\,{{\bar \theta}_{\alpha}}$, while  

\be 
\nbar{({\bar \partial}_{\alpha})}~=~-\partial^{\alpha}\ ;\qquad\overline{(\partial^{\beta})}~=~-\,{\bar \partial_{\beta}}\ .
\ee
Also, the order of the operators is interchanged; that is $\nbar{{\theta^{\alpha}}\theta^{\beta}}\,=\,{{\bar \theta}_{\beta}\,\bar\theta_{\alpha}}$, and 
$ \nbar{\partial^{\alpha}\,\partial^{\beta}}~=~{\bar \partial}_{\beta}\,{\bar \partial}_{\alpha}$.

\noindent The chiral derivatives in superspace are,  

\bea
{d^{\,{\alpha}}}=-{\partial^{\alpha}}\,-\,\frac{i}{\sqrt 2}\,{\theta^{\alpha}}\,{\partial^+}\ ;\qquad{{\bar d}_{\,{\beta}}}=\;\;\;{{\bar \partial}_{\beta}}\,+\,\frac{i}{\sqrt 2}\,{{\bar \theta}_{\beta}}\,{\partial^+}\ ,
\eea
and satisfy,

\be
\{\,{d^{\alpha}}\,,\,{{\bar d}_{\beta}}\,\}\,=\,-i\,{\sqrt 2}\,{{\delta^{\alpha}}_{\beta}}\,{\parp}\ .
\ee

\noindent The $SO(7)$ $\gamma$-matrices are defined by
\bea
\begin{split}
Tr\,(\,{\gamma^m}\,{\gamma^n}\,)\;&=\;{8}\;{\delta^{mn}}.
\end{split}
\eea
We also define,
\bea
\gamma^{mn}\,=\,\frac{1}{2}\,[\,\gamma^m\,,\,\gamma^n\,]\;,\;\;\;m\,\neq\,n.
\eea

\end{document}